\documentclass[graybox]{svmult}

\usepackage{type1cm}        % activate if the above 3 fonts are
\usepackage{makeidx}         % allows index generation
\usepackage{graphicx}        % standard LaTeX graphics tool
\usepackage{multicol}        % used for the two-column index
\usepackage[bottom]{footmisc}% places footnotes at page bottom

\usepackage{newtxtext}       % 
\usepackage{newtxmath}       % selects Times Roman as basic font

\usepackage{url}
\usepackage{hyperref}
\usepackage{bookmark}
\usepackage{color}
\usepackage{framed}

\newcommand{\R}{\mathbb R}
\newcommand{\Q}{\mathbb Q}
\newcommand{\Z}{\mathbb Z}
\newcommand{\N}{\mathbb N}
\newcommand{\K}{\mathbb K}

\newcommand{\ie}{i.e.,\ }
\newcommand{\eg}{e.g.,\ }

\newcommand{\SigmaP}{\texttt{Sigma}}
\newcommand{\MultiIntegrateP}{\texttt{Multi\-Inte\-grate}}
\newcommand{\HarmonicSumsP}{\texttt{Harmonic\-Sums}}
\newcommand{\ep}{\varepsilon}

\newcounter{mmacnt}
\def\restartmma{\setcounter{mmacnt}{0}}
\restartmma \catcode`|=\active
\def|#1|{\mathrm{#1}}
\catcode`|=12
\newenvironment{mma}{
 \par\smallskip
 \catcode`|=\active
 \parskip=0pt\parindent=0pt % locally
 \small
 \def\In##1\\{%
   \def\linebreak{\hfill\break\null\qquad}%
   \refstepcounter{mmacnt}
   \hangindent=2.0em\hangafter=0
   \leavevmode
   \llap{\tiny\sffamily In[\arabic{mmacnt}]:=\kern.5em}%
   \mathversion{bold}\footnotesize$\displaystyle##1$\normalsize
   \mathversion{normal}\par
 }%
 \def\Print##1\\{%
   \def\linebreak{\hfill\break}%
   \hangindent=2.0em\hangafter=0
   \leavevmode ##1\par}%
 \def\Out##1\\{%
   \def\linebreak{$\hfill\break\null\hfill$}%
   \kern\abovedisplayskip\par
   \hangindent=2.0em\hangafter=0
   \leavevmode
   \llap{\tiny\sffamily Out[\arabic{mmacnt}]=\kern.5em}
   \footnotesize$\displaystyle##1$\normalsize\hfill\null\par
   \kern\belowdisplayskip
 }%
 \def\Warning##1##2\\{%
   \def\linebreak{\hfill\break}%
   \hangindent=2.0em\hangafter=0
   \leavevmode
   {\scriptsize##1 : ##2}\par}%
}{%
 \par\smallskip
}

\makeindex             % used for the subject index
\begin{document}

\title*{Extensions of the AZ-algorithm and the Package MultiIntegrate}
%\titlerunning{Extensions of the AZ-algorithm}
\author{Jakob Ablinger}
\institute{Jakob Ablinger \at Johannes Kepler University Linz, Research Institute for Symbolic Computation (RISC), Altenberger Straße 69, A–4040, Linz, Austria, \email{Jakob.Ablinger@risc.jku.at}}

\maketitle

\begin{flushleft}
\vspace{-7cm}
RISC Report Series No. 21-02\\
\end{flushleft}
\vspace{6.5cm}

\abstract*{We extend the (continuous) multivariate Almkvist-Zeilberger algorithm in order to apply it for instance to special Feynman integrals emerging in renormalizable Quantum field Theories. We will consider multidimensional integrals over hyperexponential integrands and try to find closed form representations in terms of nested sums and products or iterated integrals. In addition, if we fail to compute a closed form solution in full generality, we may succeed in computing the first coefficients of the Laurent series expansions of such integrals in terms of indefinite nested sums and products or iterated integrals. In this article we present the corresponding methods and algorithms. 
Our Mathematica package \texttt{MultiIntegrate}, can be considered as an enhanced implementation of the (continuous) multivariate Almkvist Zeilberger algorithm to compute 
recurrences or differential equations for hyperexponential integrands and integrals. Together with the summation package \texttt{Sigma} and the package \texttt{HarmonicSums} our package provides methods to compute closed form representations (or coefficients of the Laurent series expansions) of multidimensional integrals over hyperexponential integrands in terms of nested sums or iterated integrals.}

\abstract{We extend the (continuous) multivariate Almkvist-Zeilberger algorithm in order to apply it for instance to special Feynman integrals emerging in renormalizable Quantum field Theories. We will consider multidimensional integrals over hyperexponential integrals and try to find closed form representations in terms of nested sums and products or iterated integrals. In addition, if we fail to compute a closed form solution in full generality, we may succeed in computing the first coefficients of the Laurent series expansions of such integrals in terms of indefinite nested sums and products or iterated integrals. In this article we present the corresponding methods and algorithms. 
Our Mathematica package \texttt{MultiIntegrate}, can be considered as an enhanced implementation of the (continuous) multivariate Almkvist Zeilberger algorithm to compute 
recurrences or differential equations for hyperexponential integrands and integrals. Together with the summation package \texttt{Sigma} and the package \texttt{HarmonicSums} our package provides methods to compute closed form representations (or coefficients of the Laurent series expansions) of multidimensional integrals over hyperexponential integrands in terms of nested sums or iterated integrals.}

\section{Introduction}
The Almkvist-Zeilberger was first formulated by Apagodu and Zeilberger~\cite{AlmZeil,mohammed05} and has later been refined and generalized~\cite{AblingerDiss,chen12b,chen12c,chen14a,koutschan10b}. It attracted attention in renormalizable Quantum Field Theory in the frame of the calculation of Feynman integrals.

In the following we briefly want to summarize the structure of those integrals. 
The very general class of Feynman integrals which are, for instance, considered in~\cite{Bluemlein2011} are of relevance for many physical processes at high energy colliders, 
such as the Large Hadron Collider and others.
The considered integrals are two--point
Feynman integrals in $D$-dimensional Minkowski space with one time- and $(D-1)$
Euclidean space dimensions, $\ep = D - 4$ and $\ep \in {\mathbb R}$ with
$|\ep| \ll 1$ of the
following structure:
%--------------------------------------------------------------------------
\begin{eqnarray}
{\cal I}(\ep,N,p) = \int \frac{d^D p_1}{(2\pi)^D} \ldots \int \frac{d^D
p_k}{(2\pi)^D}
\frac{{\cal N}(p_1, \ldots p_k; p; m_1 \ldots
m_k;
\Delta, N)}{(-p_1^2 + m_1^2)^{l_1} \ldots (-p_k^2 + m_k^2)^{l_k}}
\prod_V \delta_V~.
\label{eq:A7}
\end{eqnarray}
%---------------------------------------------------------------------------
They can be shown to obey difference equations with respect to $N,$ see, \eg \cite{BKKS}.
In (\ref{eq:A7}) the external momentum $p$ and the loop momenta $p_i$ denote $D$-dimensional
vectors, $m_i > 0 , m_i \in
{\mathbb R}$ are scalars
(masses),
$m_i \in \{0, M\}$,
$k, l_i \in {\mathbb N}$, $k \geq 2,l_i \geq 1$, and $\Delta$ is a light-like
$D$-vector,
$\Delta.\Delta = 0$. The numerator function ${\cal N}$ is a polynomial in the
scalar products $p.p_i,~p_i.p_k$ and of monomials $(\Delta.p_{(i)})^{n_i}$, $n_i \in {\mathbb N},
n_i \geq 0$. $N \in {\mathbb N}$ denotes the spin of a local operator stemming from
the light cone expansion, see, e.g., \cite{Frishman1971} and references therein, which
contributes to the numerator function ${\cal N}$ with a polynomial in $\Delta.p_i$ of
maximal degree $N$, cf. \cite{Bierenbaum2009}. Furthermore it is assumed for simplicity that only
one of the loops is formed of massive lines. The $\delta_V$ occurring in (\ref{eq:A7}) are shortcuts for Dirac delta distributions in $D$ dimensions 
$\delta_V = \delta^{(D)}\left(\sum_{l=1}^k a_{V,l}p_l\right), a_{V,l}\in\Q.$\\
These integrals are mathematically well defined and in \cite{Bluemlein2011} it is shown how they can be mapped onto
integrals on the $m$-dimensional unit cube with the following structure:
\begin{eqnarray}
{\cal I}(\ep,N) = C(\ep, N, M) \int_0^1 dy_1 \ldots \int_0^1 dy_m
\frac{\sum_{i=1}^k \prod_{{l}=1}^{r_i}
[P_{i,l}(y)]^{\alpha_{i,l}(\ep,N)}}{[Q(y)]^{\beta(\ep)}}~,
\label{eq:A9}
\end{eqnarray}
%---------------------------------------------------------------------------
with $k\in\N$, $r_1,\dots,r_k\in\N$ and
where $\beta(\ep)$ is  given by a rational function in $\ep$, i.e., $\beta(\ep)\in\R(\ep)$, and similarly
$\alpha_{i,l}(\ep,N) = n_{i,l} N + \overline{\alpha}_{i,l}$ for some $n_{i,l} \in \{0,1\}$ and $\overline{\alpha}_{i,l}\in\R(\ep)$, see also \cite{Bogner2010}
in the case no local operator insertions are present.
$C(\ep, N, M)$ is a factor, which depends on the dimensional parameter $\ep$,
the integer parameter $N$ and the mass $M$.
$P_i(y), Q(y)$ are polynomials in the remaining Feynman parameters $y=(y_1,\dots,y_m)$ written in multi-index notation.
In (\ref{eq:A9}) all terms
which stem from local operator insertions were geometrically resumed; see~\cite{Bierenbaum2009}.
In \cite{Bluemlein2011} it was already mentioned that after splitting the integral (\ref{eq:A9}), the integrands fit into the input class of the multivariate Almkvist-Zeilberger algorithm. Hence, if 
the split integrals are properly defined, they obey homogeneous recurrence relations in $N$ due to the theorems in \cite{AlmZeil}. In \cite{Bluemlein2011} the integrals of (\ref{eq:A9}) are transformed
further to a multi-sum representation, while in this article (and also in \cite{AblingerDiss,Ablinger:2014lka,Ablinger2016}) we want to tackle them directly by looking on integrals of the form
\begin{align}
\label{AZhypexpint}
{\cal I}(n)={\cal I}(\ep,n) = \int_{u_d}^{o_d} \dots\int_{u_1}^{o_1}F(n;x_1, \dots, x_d;\ep)\; dx_1 \cdots dx_d,
\end{align}
with $d,n \in \N$, $F(n;x_1, \dots, x_d;\ep)$ a hyperexponential term, $\ep>0$ a real parameter and $u_i,o_i \in \R\cup \{-\infty,\infty\}.$\\
In \cite{AblingerDiss} we only considered a discrete variable $n$ but here we will also consider continuous variables, \ie we will also deal with integrals of the form
\begin{align}
\label{cAZhypexpint}
{\cal I}(x)={\cal I}(\ep,x) = \int_{u_d}^{o_d} \dots\int_{u_1}^{o_1}F(x;x_1, \dots, x_d;\ep)\; dx_1 \cdots dx_d,
\end{align}
with $d \in \N$, $x \in \R $, $F(x;x_1, \dots, x_d;\ep)$ a hyperexponential term, $\ep>0$ a real parameter and $u_i,o_i \in \R\cup \{-\infty,\infty\}.$
We will use our package \MultiIntegrateP\footnote{\label{download}The {\tt Mathematica} packages {\tt MultiIntegrate},  {\tt HarmonicSums}, {\tt Sigma} and {\tt Evaluate\-MultiSums}  can be downloaded at \url{https://risc.jku.at/software}}\cite{AblingerDiss,Ablinger2016} that can be considered as an enhanced implementation of the multivariate Almkvist Zeilberger algorithm to compute recurrences/differential equations for the integrands and integrals. For solving recurrences \MultiIntegrateP\ relies on the solver implemented in the packages \SigmaP\footref{download} \cite{Schneider2007,Schneider2013,Schneider2014} and \texttt{EvaluateMultiSums}\footref{download} \cite{Schneider2013,Schneider2014}, while for solving differential equation it relies on the solver implemented in the package \HarmonicSumsP\footref{download}~\cite{Ablinger:2011te,AblingerDiss,Ablinger:2013cf,HarmonicSums,Ablinger:2014bra}.

Throughout this article $\K$ denotes a field with $\Q\subseteq\K$ (\eg $\K =\Q(\ep)$ forms a rational function field) in which the usual operations can be computed.

The remainder of this article is structured in two main sections. In Section~\ref{Sec:AZ} we will recall the multi-variate Almkvist-Zeilberger algorithm and its modifications as presented in~\cite{AblingerDiss} to solve integrals of the form~\eqref{AZhypexpint}, while in Section~\ref{Sec:cAZ} we will present a method based on the continuous Almkvist-Zeilberger algorithm to solve integrals of the form~\eqref{cAZhypexpint}. The reason for Section~\ref{Sec:AZ} is its similarity to the continuous case in Section~\ref{Sec:cAZ}, in addition it allows us to present a complete picture of the functionality of the package \MultiIntegrateP. However, in order to keep Section~\ref{Sec:AZ} short, we omit examples in this section and refer to~\cite{AblingerDiss,Ablinger:2014lka,Ablinger2016} for further illustrations.   
\section{A fine-tuned multi-variate Almkvist-Zeilberger algorithm}
\label{Sec:AZ}
In this section we will recall a method (presented in~\cite{AblingerDiss} and~\cite{Ablinger2016}) to compute integrals of the form \eqref{AZhypexpint}, that is based on slight modifications of the multi-variate Almkvist-Zeilberger algorithm~\cite{AlmZeil} and implemented in the package \MultiIntegrateP. The method relies on finding and solving recurrences.\\
In general, consider the integrand
%----------------------------------------------------------------------------------------------------
\begin{align}
F(n;x_1, \dots , x_d)=P(n;x_1, \dots, x_d) \cdot H(n;x_1, \dots, x_d), \label{mAZintegrand}
\end{align}
%----------------------------------------------------------------------------------------------------
with a multivariate polynomial $P(n;x_1, \dots, x_d) \in \K[n,x_1, \dots, x_d]$ and
%----------------------------------------------------------------------------------------------------
\begin{align*}
H(n;x_1, \dots, x_d)=
e^{\frac{a(x_1, \dots,x_d)}{b(x_1, \dots, x_d)}} \cdot
\left ( \prod_{p=1}^P {S_p(x_1, \dots, x_d)}^{\alpha_p} \right )\cdot
\left ( { \frac{s(x_1, \dots, x_d)}{t(x_1, \dots, x_d)} } \right )^n,
\end{align*}
%----------------------------------------------------------------------------------------------------
where $a(x_1, \dots, x_d)$ and $b(x_1, \dots, x_d)\neq 0$,
$s(x_1, \dots, x_d)$ and $t(x_1, \dots, x_d)\neq 0$ and
$S_p(x_1, \dots, x_d) \in \K[x_1, \dots, x_d]$, and $\alpha_p\in \K$. Such integrands have the property that the logarithmic derivatives are rational, $\ie$
$$
{ \frac{D_{x_i} \overline{H}(n;x_1, \dots, x_d)}{\overline{H}(n;x_1, \dots, x_d)} }={\frac{q_i(x_1, \dots, x_d)}{r_i(x_1, \dots, x_d)}}
$$
for some $q_i(n,x_1, \dots, x_d),r_i(n,x_1, \dots, x_d)\in\K[n,x_1,\dots,x_d]$ and are called \textit{hyperexponential} in $x_i$.
Note that this class of integrands 
covers a big class of Feynman integrals (by choosing the rational function field $\K=\Q(\ep)$) that contains at most one mass~\cite{Bluemlein2011,Weinzierl:13}.\\
Then due to~\cite{AlmZeil} there exists a non-negative integer $L$, 
there exist $e_0(n), \dots , e_L(n)\in\K[n]$ (or equivalently from $\K(n)$),
{\it not all zero}, and  there also exist $R_i(n;x_1, \dots, x_d)\in\K(n,x_1, \dots, x_d)$
%($i=1, \dots ,d$) 
such that
%----------------------------------------------------------------------------------------------------
\begin{align}
G_i(n;x_1, \dots, x_d):=R_i(n;x_1, \dots,x_d)F(n;x_1, \dots, x_d)  \label{Gs}
\end{align}
%----------------------------------------------------------------------------------------------------
satisfy the integrand recurrence
%----------------------------------------------------------------------------------------------------
\begin{align}
\sum_{i=0}^L e_i(n) F(n+i;x_1, \dots, x_d)= \sum_{i=1}^d D_{x_i} G_i(n;x_1, \dots, x_d), \label{mAZrec}
\end{align}
%----------------------------------------------------------------------------------------------------
where $D_{x_i}$ stands for the derivative w.r.t $x_i$.
\subsection{The general method} 
The proof of the existence, and in particular a method to compute such an integrand recurrence~\eqref{mAZrec}, 
is based on the following observation~\cite{AlmZeil}.
Fix a non-negative integer $L$ (with the role given above) and define
%----------------------------------------------------------------------------------------------------
\begin{align*}
\overline{H} (n;x_1, \dots, x_d)
    :=e^{\frac{a(x_1, \dots,x_d)}{b(x_1, \dots, x_d)}} \cdot\left (\prod_{p=1}^P {S_p(x_1, \dots, x_d)}^{\alpha_p} \right )\cdot{ \frac{s(x_1, \dots, x_d)^n}{t(x_1, \dots, x_d)^{n+L}}},
\end{align*}
%---------------------------------------------------------------------------------------------------------
Then we have 
$$
 \sum_{i=0}^{L} e_i(n) F(n+i;x_1, \dots, x_d)= h(x_1,\ldots,x_d)\overline{H} (n;x_1, \dots, x_d),
$$
 where $h(x_1,\ldots,x_d)$ is a polynomial \ie
$$
h(x_1,\ldots,x_d):=\sum_{i=1}^{L}e_i(n)P(n+i,x_1,\ldots,x_d)\frac{s(x_1,\ldots,x_d)^i}{t(x_1,\ldots,x_d)^{i-L}}.
$$
and, by construction, the logarithmic derivatives of $\overline{H} (n;x_1, \dots, x_d)$ are a rational functions in the $x_i$, i.e., we have that
$$
{ \frac{D_{x_i} \overline{H}(n;x_1, \dots, x_d)}{\overline{H}(n;x_1, \dots, x_d)} }={\frac{q_i(x_1, \dots, x_d)}{r_i(x_1, \dots, x_d)}}
$$
%----------------------------------------------------------------------------------------------------
for explicitly given $q_i(n,x_1, \dots, x_d),r_i(n,x_1, \dots, x_d)\in\K[n,x_1,\dots,x_d]$.\\ 
For $i=1, \dots, d$ we make the general ansatz
%----------------------------------------------------------------------------------------------------
\begin{align}
G_i(n;x_1, \dots, x_d)=\overline{H}(n;x_1, \dots, x_d) \cdot r_i(n,x_1,\dots, x_d ) \cdot X_i(n; x_1, \dots, x_d). \label{mAZansatz}
\end{align}
%----------------------------------------------------------------------------------------------------
Then it turns out that for $L$ chosen sufficiently large\footnote{There exist upper bounds for a particular input. But usually, these bounds are too high and one tries smaller values.} there exist polynomials
%----------------------------------------------------------------------------------------------------
$X_i(n; x_1, \dots , x_d)\in \K[n][x_1, \dots, x_d]$ with $1\leq i\leq L$ and polynomials $e_i(n)\in\K[n]$ (not all zero) such that~\eqref{mAZrec} holds. Motivated by this fact, one searches for these unknowns $X_i$ and $e_i$ as follows. 
Note that the ansatz~(\ref{mAZrec}) is equivalent to (see~\cite{AlmZeil})
%----------------------------------------------------------------------------------------------------
\begin{align}
&\sum_{i=1}^d [D_{x_i}r_i(x_1, \dots, x_d)+q_i(x_1, \dots, x_d)] \cdot X_i(n;x_1, \dots, x_d)\nonumber\\
&\hspace{1cm}+r_i(x_1, \dots, x_d)\cdot D_{x_i}X_i(n;x_1, \dots, x_d)\nonumber \\
&=\sum_{i=0}^L e_i(n)\,P(n;x_1,\dots,x_d)\,s(x_1,\dots,x_d)^i\,t(x_1,\dots,x_d)^{L-i}. \label{mAZrec2}
\end{align}
%----------------------------------------------------------------------------------------------------
We choose appropriate degree bounds w.r.t.\ the $x_1,\dots,x_d$ for the $X_i$ ($1\leq i\leq d$) and plug the polynomials 
with unknown coefficients from $\K[n]$ (from $\K(n)$) into~\eqref{mAZrec2}. By coefficient comparison this yields 
a linear system in $\K(n)$ with the unknowns $e_i(n)$ and the unknown coefficients of the polynomials $X_i$. Finally, we can seek for a non-trivial solution for~\eqref{mAZrec2} and thus for~\eqref{mAZrec}. To optimize the search for a non-trivial solution we make use of homomorphic image computations in our implementation. More precisely, we plug in some concrete integers for the parameters and reduce all integer coefficients modulo a prime. If there is no solution in the homomorphic setting, there is no solution in the general setting. By choosing these values sufficiently generically we can also minimize the risk of obtaining a homomorphic solution that does not extend to a general solution. In the end, we clear 
denominators in $n$ such that the $e_i(n)$ turn to polynomials.\\
If $F(n;\dots,x_{i-1},u_i,x_{i+1},\dots)=0$ and $F(n;\dots,x_{i-1},o_i,x_{i+1},\dots)=0$ 
%(and hence $G(n;\dots,x_{i-1},u_i,x_{i+1},\dots)=0$ and $G(n;\dots,x_{i-1},o_i,x_{i+1},\dots)=0$ )
then
%----------------------------------------------------------------------------------------------------
$$
{\cal I}(n):=\int_{u_d}^{o_d} \dots\int_{u_1}^{o_1}F(n;x_1, \dots, x_d) dx_1 \dots dx_d,
$$
%----------------------------------------------------------------------------------------------------
satisfies the homogeneous linear recurrence equation with polynomial coefficients 
%----------------------------------------------------------------------------------------------------
\begin{align}\label{AZlinrec}
\sum_{i=0}^L e_i(n) {\cal I}(n+i)= 0.
\end{align}
The general method now is straightforward: Given an integrand of the form~(\ref{mAZintegrand}), we can set $L=0,$ look for degree bounds for $X_i(x_1, \dots , x_d)$ and try to find a 
solution of~(\ref{AZlinrec}) by coefficient comparison. If we do not find a solution of~(\ref{AZlinrec}) with not all $e_i(n)$'s equal to zero (we stop the calculation if the homomorphic image check fails), we increase $L$ by one, look for new degree bounds 
for $X_i(x_1, \dots , x_d)$ and try again to find a solution of~(\ref{AZlinrec}). Again, if we do not find a solution  with not all $e_i(n)$'s equal to zero, we increase $L$ by one and repeat 
the process.
\begin{svgraybox}
The discrete multiple Almkvist-Zeilberger algorithm is implemented in the command \texttt{mAZ} of \MultiIntegrateP.
\end{svgraybox}
Once we found a recurrence we exploit algorithms from~\cite{Abramov1994,Petkovsek1992,Schneider2001,Schneider2006} which can constructively decide if a solution with certain initial values 
is expressible in terms of indefinite nested products and sums. This covers harmonic sums \cite{Bluemlein1999,Vermaseren1998}, S-sums \cite{Ablinger:2013cf,Moch2002}, cyclotomic sums \cite{Ablinger:2011te} and binomial sums \cite{Ablinger:2014bra,Kalmykov:2000qe} as special cases. In our implementation we make use of the algorithms implemented in the summation package~\SigmaP. For details on which solutions can be found using \SigmaP,  we refer to \cite{Bluemlein2011}.
\subsection{Dealing with non-standard boundary conditions}
Unfortunately, in many cases the integrand (\ref{mAZintegrand}) does not vanish at the integration bounds and we end up in a linear recurrence with a non-trivial inhomogeneous part which can be written as a linear combination of integrals with at least one integral operator less.
In the following we will deal with non-standard boundary conditions in two different ways, see \cite {AblingerDiss}.
\subsubsection{Dealing with inhomogeneous recurrences}
\label{inhomansatz}
In \cite{AblingerDiss} a method that deals with the inhomogeneous recurrence similar to \cite{Bluemlein2011} can be found. It gives rise to a recursive method. To be more precise, we consider the integral
$$
{\cal I}(n):=\int_{u_d}^{o_d} \cdots\int_{u_1}^{o_1}F(n;x_1, \dots, x_d) dx_1 \dots dx_d.
$$
Suppose that we found
\begin{equation}
\sum_{i=0}^L e_i(n) F(n+i;x_1, \dots, x_d)= \sum_{i=1}^d D_{x_i} G_i(n;x_1, \dots, x_d)
\end{equation}
where at least one $G_i(n;x_1, \dots, x_d)$ does not vanish at the integration limits. By integration with respect to $x_1,\ldots,x_d$ we can deduce that ${\cal I}(n)$ satisfies the inhomogeneous linear recurrence 
equation
\begin{align*}
&\sum_{i=0}^L e_i(n) {\cal I}(n+i)=\\
&\hspace{1cm}\sum_{i=1}^d \int_{u_d}^{o_d} \cdots \int_{u_{i-1}}^{o_{i-1}}\int_{u_{i+1}}^{o_{i+1}}\cdots \int_{u_1}^{o_1}O_i(n)dx_1\dots dx_{i-1}dx_{i+1}\dots dx_d\\
&\hspace{1cm}-\sum_{i=1}^d \int_{u_d}^{o_d} \cdots \int_{u_{i-1}}^{o_{i-1}}\int_{u_{i+1}}^{o_{i+1}}\cdots \int_{u_1}^{o_1}U_i(n)dx_1\dots dx_{i-1}dx_{i+1}\dots dx_d
\end{align*}
with
\begin{align*}
U_i(n)&:=G_i(n;x_1,\dots,x_{i-1},o_i,x_{i+1} \dots, x_d)\\
O_i(n)&:=G_i(n;x_1,\dots,x_{i-1},u_i,x_{i+1} \dots, x_d).
\end{align*}
Note that the inhomogeneous part of the above recurrence equation is a sum of $2\cdot d$ integrals of dimension $d-1,$ which fit again into the input class of the multiple Almkvist-Zeilberger algorithm. 
Hence we can apply the algorithms to the $2\cdot d$ integrals recursively until we arrive at the base case of one-dimensional integrals for which we have to solve an inhomogeneous linear recurrence relation 
where the inhomogeneous part is free of integrals. Given the solutions for the one-dimensional integrals we can step by step find the solutions of higher dimensional integrals until we finally find the 
solution for ${\cal I}(n)$ by solving again an inhomogeneous linear recurrence equation and combining it with the initial values. Note that we have to calculate initial values with respect to $n$ for all the integrals 
arising in this process.

Summarizing, with these algorithms we use the following strategy (note that we assume that we are able to compute the initial values for the arising integrals); compare \cite{AblingerDiss,Bluemlein2011}:
\begin{programcode}{Divide and conquer strategy}
\vspace{-0.6cm}
\begin{enumerate}
\item BASE CASE: If ${\cal I}(n)$ has no integration quantifiers, return ${\cal I}(n).$

\item DIVIDE: As worked out above, compute a recurrence relation
\begin{equation}\label{Equ:IntRec}
a_0(n){\cal I}(n)+\dots+a_d(n){\cal I}(n+d)=h(n)
\end{equation}
with polynomial coefficients
$a_i(n)\in\K[n]$, $a_m(n)\neq0$ and the right side $h(n)$ containing a linear
combination of hyperexponential multi-integrals each with less than $d$ integration
quantifiers.
\item CONQUER: Apply the strategy recursively to the simpler integrals in
$h(n)$. This results in an indefinite nested product-sum expressions $\tilde{h}(n)$ with 
\begin{equation}\label{Equ:hsol}
\tilde{h}(n)=h(n),\quad \forall n\geq \delta \text{ for some } \delta \in \N.
\end{equation}
If the method fails to find the $\tilde{h}(n)$ in terms of indefinite
nested product-sum expressions, STOP.
\item COMBINE: Given~\eqref{Equ:IntRec} with~\eqref{Equ:hsol},
compute, if possible, $\tilde{{\cal I}}(n)$ in terms of nested product-sum expressions such that
\begin{equation}
\tilde{{\cal I}}(n)={\cal I}(n),\quad \forall n\geq \delta \text{ for some } \delta \in \N.
\end{equation}
by solving the recurrence.
\end{enumerate}
\end{programcode}
\begin{svgraybox}
This divide and conquer strategy is implemented in the command \texttt{mAZ\-Inte\-grate} of \MultiIntegrateP. 
\end{svgraybox}
\begin{remark}
We remark that this approach works nicely, if the initial values of the integrals in 
the inhomogeneous part can be calculated efficiently. Further details on this approach are given 
in~\cite{AblingerDiss,LL12:Technolgy}. We remark further that similar approaches have been explored in~\cite{Bluemlein2011,Round2018} and~\cite{LL12:Technolgy} based on~\cite{Wegschaider,WZ}
and~\cite{NewSigmaApproach}, respectively, in order to derive recurrences for hypergeometric
multi-sums.
\end{remark}
\subsubsection{Adapting the ansatz to find homogeneous recurrences}
\label{homansatz}
In order to avoid the difficulties of inhomogeneous recurrences we adapt the ansatz.
Namely, we can always obtain a homogeneous recurrence of the form~\eqref{AZlinrec}
by changing (\ref{mAZansatz}) to
%----------------------------------------------------------------------------------------------------
\begin{align}
&G_i(n;x_1, \dots, x_d)=\nonumber\\&\hspace{1cm}\overline{H}(n;x_1, \dots, x_d) \cdot r_i(x_1,\dots, x_d ) \cdot X_i(x_1, \dots, x_d)(x_i-u_i)(x_i-o_i), \label{mAZansatz2}
\end{align}
%----------------------------------------------------------------------------------------------------
\ie the $G_i$ are forced to vanish at the integration bounds. 
Then with this Ansatz (\ref{mAZrec2}) the underlying linear system turns into
%----------------------------------------------------------------------------------------------------
\begin{eqnarray}
&&\sum_{i=1}^d [D_{x_i}r_i(x_1, \dots, x_d)+q_i(x_1, \dots, x_d)] \cdot X_i(x_1, \dots, x_d)(x_i-u_i)(x_i-o_i) \nonumber \\
&&\hspace{1cm}+r_i(x_1, \dots, x_d)\cdot D_{x_i}X_i(x_1, \dots, x_d)(x_i-u_i)(x_i-o_i)\nonumber\\
&&\hspace{1cm}=\sum_{i=0}^L e_i(N)\,P(N;x_1,\dots,x_d)\,s(x_1,\dots,x_d)^i\,t(x_1,\dots,x_d)^{L-i}.\label{mAZrec3}
\end{eqnarray}
%----------------------------------------------------------------------------------------------------
The general method now is straightforward: Given an integrand of the form~(\ref{mAZintegrand}), we can set $L=0,$ look for degree bounds for $X_i(x_1, \dots , x_d)$ and try to find a 
solution of~(\ref{mAZrec3}) by coefficient comparison. If we do not find a solution of~(\ref{mAZrec3}) with not all $e_i(n)$'s equal to zero, we increase $L$ by one, look for new degree bounds 
for $X_i(x_1, \dots , x_d)$ and try again to find a solution of~(\ref{mAZrec3}). Again, if we do not find a solution  with not all $e_i(n)$'s equal to zero, we increase $L$ by one and repeat 
the process.\\
Once we found a recurrence we can use the recurrence solver implemented in the summation package~\SigmaP\ to try to solve it.
\begin{svgraybox}
This strategy is implemented in the command \texttt{mAZ\-Direct\-Inte\-grate} of \MultiIntegrateP.
\end{svgraybox}
\begin{remark}
The advantage of this approach is, that we do not have to deal with integrals (and initial conditions) recursively, since the recurrence is homogenous, however the additional conditions on the ansatz might increase the order of the recurrence drastically. In particluar, the routine is more robust: no abortion can occur due to problematic integral arising from the reccursion.
\end{remark}
\subsection{Computing series expansions of the integrals}
\label{Sec:AZseriesexp}
Due to time and memory limitations, not finding all solutions of the recurrences or due to missing initial values (in full generality) we might fail to process certain integrals using the methods described in the previous subsection. Therefore, inspired 
by \cite{Bluemlein2011}, a method which computes $\ep$-expansions of integrals of the form (\ref{AZhypexpint}) was developed in \cite{AblingerDiss}. In the following we recall this method. We assume that the
integral ${\cal I}(\ep,n)$ from (\ref{AZhypexpint}) has a Laurent expansion in $\ep$ for each $n\in\N$ with $n\geq\lambda$ for some $\lambda\in\N$ and 
thus it is an analytic function in $\ep$ throughout an annular region centered by $0$ where the pole at $\ep=0$ has some order~$K\in \Z$. Hence we can write it in the form
\begin{equation}
 {\cal I}(\ep,n) = \sum_{k=-K}^{\infty} \ep^k I_k(n).
\end{equation}
In the following we try to find the first coefficients $I_t(n),I_{t+1}(n),\ldots,I_{u}(n)$ in terms of indefinite nested product-sum expressions of the expansion
\begin{equation}\label{Equ:FExp2}
{\cal I}(\ep, n) = I_t(n)\ep^t+I_{t+1}(n)\ep^{t+1}+I_{t+2}(n)\ep^{t+2}+\dots
\end{equation}
with $t=-K \in \Z$.
We start by computing a recurrence for  ${\cal I}(\ep,n)$ in the form
\begin{multline}\label{Equ:ExpansionEquMod}
a_0(\ep,n)J(\ep,n)+a_1(\ep,n)J(\ep,n+1)+\dots+a_d(\ep,n)J(\ep,n+d)\\
=h_{-K}(n)\ep^{-K}+h_{-K+1}(n)\ep^{-K+1}+\dots+h_{u}(n)\ep^u+\dots
\end{multline}
In order to accomplish this task, we can use the methods presented in the previous section.
Given the recurrence we exploit an algorithm from~\cite{Bluemlein2011} which can constructively decide if a formal Laurent series solution with certain initial values is expressible (up to a certain order) 
in terms of indefinite nested products and sums. This algorithm is implemented in the package \SigmaP\ and can be summarized as follows (see~\cite{Bluemlein2011} and compare \cite{AblingerDiss,Ablinger2016}).

Suppose we are given the linear recurrence \eqref{Equ:ExpansionEquMod}
of order $d$ where the $a_i(\ep,n)$ are polynomials in $n$ and $\ep$ and where the inhomogeneous part can be expanded in $\ep$ up to order $u$.
Consider a function which has a Laurent series expansion
\begin{equation}
{\cal I}(\ep,n)=F_{t}(n)\ep^t+F_{t+1}(n)\ep^{t+1}+\dots
\end{equation}
and which is a solution of the given recurrence for all $n\geq n_0$ for some $n_0\in\N$. Then together with the
$d$ initial values $F_{j}(n_0),\dots,F_j(n_0+d-1)$
with $t\leq j\leq u$, all values $F_t(n),\dots,F_u(n)$ with $n\geq n_0$ can be computed provided that the values 
$h_i(n)$ for all $i$ with $t\leq i\leq u$ and all integers $n$ with $n\geq n_0$ can be computed. In addition, if the $h_t(n),\dots,h_u(n)$ are given explicitly in terms of indefinite nested product-sum expressions, there is an algorithm which decides constructively if the $F_{t}(n),\dots,F_u(n)$ can be given in terms of indefinite nested product-sum expressions.

Having such a Laurent series recurrence solver in hand we can combine it with the methods from the previous sections.
Let ${\cal I}(\ep,n)$ be a multi-integral of the form~\eqref{AZhypexpint} and assume that ${\cal I}(\ep,n)$ has a series expansion~\eqref{Equ:FExp2} for all $n\geq\lambda$ for 
some $\lambda\in\N$. If we succeed in finding a homogeneous differential equation, for instance by using the method form Section~\ref{homansatz} we can directly apply the Laurent series recurrence solver, supposing that we can handle the initial values. This has been exploited in the frame of \cite{Ablinger2016}.
\begin{svgraybox}
This strategy is implemented in the command \texttt{mAZ\-Expanded\-Direct\-Inte\-grate} of \MultiIntegrateP.
\end{svgraybox}
\noindent Of course we can again think of a recursive method to compute the first coefficients (compare \cite{AblingerDiss,Bluemlein2011}), say $F_t(n),\dots,F_u(n)$ of~\eqref{Equ:FExp2}. Note that we have the same advantages and disadvantages as mentioned for the recursive method in Section \ref{inhomansatz}, but if we assume that we can handle the initial values we can use the following strategy.
\begin{programcode}{Divide and conquer strategy}
\vspace{-0.6cm}
\begin{enumerate}
\item BASE CASE: If ${\cal I}(\ep, n)$ has no integration quantifiers, compute the expansion by standard methods.
\item DIVIDE: As worked out before, compute a recurrence relation
\begin{equation}\label{Equ:IntRecurrence}
a_0(\ep,n){\cal I}(\ep, n)+\dots+a_d(\ep,n){\cal I}(\ep, n+d)=h(\ep,n)
\end{equation}
with polynomial coefficients
$a_i(\ep,n)\in\K[\ep,n]$, $a_m(\ep,n)\neq0$ and the right side $h(\ep,n)$ containing a linear
combination of hyperexponential multi-integrals each with less than $d$ integration
quantifiers.
\item CONQUER: Apply the strategy recursively to the simpler integrals in
$h(\ep,n)$. This results in an expansion of the form
\begin{equation}\label{Equ:hExpansion}
h(\ep,
n)=h_t(n)+h_1(n)\ep+\dots+h_u(n)\ep^u+O(\ep^{u+1});
\end{equation}
if the method fails to find the $h_t(n),\dots,h_u(n)$ in terms of indefinite
nested product-sum expressions, STOP.
\item COMBINE: Given~\eqref{Equ:IntRecurrence} with ~\eqref{Equ:hExpansion},
compute, if possible, the $F_t(n),\dots,F_u(n)$ of~\eqref{Equ:FExp2} in terms of
nested product-sum expressions by using \SigmaP.
\end{enumerate}
\end{programcode}
\begin{svgraybox}
This divide and conquer strategy is implemented in the command \texttt{mAZ\-Expanded\-Inte\-grate} of \MultiIntegrateP.
\end{svgraybox}
\section{A fine-tuned continuous multi-variate Almkvist-Zeilberger algorithm}
\label{Sec:cAZ}
In this section we present a method to compute integrals of the form \eqref{cAZhypexpint}, that is based on slight modifications of the continuous multi-variate Almkvist-Zeilberger algorithm~\cite{AlmZeil} and implemented in the package \MultiIntegrateP. Unlike in the discrete case, this method relies on finding and solving differential equations.
In general, consider the hyperexponential integrand
%----------------------------------------------------------------------------------------------------
\begin{align}
F(x;x_1, \dots , x_d)=P(x;x_1, \dots, x_d) \cdot H(x;x_1, \dots, x_d), \label{cmAZintegrand}
\end{align}
%----------------------------------------------------------------------------------------------------
with a multivariate polynomial $P(x;x_1, \dots, x_d) \in \K[x,x_1, \dots, x_d]$ and
%----------------------------------------------------------------------------------------------------
\begin{align*}
H(x;x_1, \dots, x_d)=
e^{\frac{a(x,x_1, \dots,x_d)}{b(x,x_1, \dots, x_d)}} \cdot
\left ( \prod_{p=1}^P {S_p(x,x_1, \dots, x_d)}^{\alpha_p} \right ),
\end{align*}
%----------------------------------------------------------------------------------------------------
where $a(x,x_1, \dots, x_d),b(x,x_1, \dots, x_d)$ and
$S_p(x,x_1, \dots, x_d) \in \K[x,x_1, \dots, x_d]$, with $b(x,x_1, \dots, x_d) \neq 0$,  and $\alpha_p\in \K$.
Then due to~\cite{AlmZeil} there exists a non-negative integer~$L$, 
there exist $e_0(x),e_1(x), \dots , e_L(x)\in\K[x]$ (or equivalently from $\K(x)$),
{\it not all zero}, and  there also exist $R_i(x;x_1, \dots, x_d)\in\K(x,x_1, \dots, x_d)$
%($i=1, \dots ,d$) 
such that
%----------------------------------------------------------------------------------------------------
\begin{align}
G_i(x;x_1, \dots, x_d):=R_i(x;x_1, \dots,x_d)F(x;x_1, \dots, x_d)  \label{cGs}
\end{align}
%----------------------------------------------------------------------------------------------------
satisfy the integrand differential equation
%----------------------------------------------------------------------------------------------------
\begin{align}
\sum_{i=0}^L e_i(x) D_{x}^i F(x;x_1, \dots, x_d)= \sum_{i=1}^d D_{x_i} G_i(x;x_1, \dots, x_d). \label{cmAZdiff}
\end{align}
%----------------------------------------------------------------------------------------------------
%
%
\subsection{The general method} 
The proof of the existence, and in particular a method to compute such a differential equation~\eqref{cmAZdiff}, 
is based on the following observation~\cite{AlmZeil}.
Fix a non-negative integer~$L$ (with the role given above), define
%----------------------------------------------------------------------------------------------------
\begin{align*}
\overline{H} (x;x_1, \dots, x_d)
    :=\frac{e^{\frac{a(x,x_1, \dots,x_d)}{b(x,x_1, \dots, x_d)}}}{b(x,x_1, \dots,x_d)^{2 L}} \cdot\left (\prod_{p=1}^P {S_p(x,x_1, \dots, x_d)}^{\alpha_p} \right ),
\end{align*}
%----------------------------------------------------------------------------------------------------
Then we have 
$$
 \sum_{i=0}^{L} e_i(x) D_{x}^i F(x;x_1, \dots, x_d)= h(x,x_1,\ldots,x_d)\overline{H} (x;x_1, \dots, x_d).
$$
for some polynomial $h(x,x_1,\ldots,x_d)$ that can be determined and,  by construction, the logarithmic derivatives of $\overline{H} (x;x_1, \dots, x_d)$ are rational functions in the $x_i$, i.e., we have that
$$
{ \frac{D_{x_i} \overline{H}(x;x_1, \dots, x_d)}{\overline{H}(x;x_1, \dots, x_d)} }={\frac{q_i(x,x_1, \dots, x_d)}{r_i(x,x_1, \dots, x_d)}}
$$
%----------------------------------------------------------------------------------------------------
for explicitly given $q_i(x,x_1, \dots, x_d),r_i(x,x_1, \dots, x_d)\in\K[x,x_1,\dots,x_d]$.\\
For $i=1, \dots, d$ we make the general ansatz
%----------------------------------------------------------------------------------------------------
\begin{align}
G_i(x;x_1, \dots, x_d)=\overline{H}(x;x_1, \dots, x_d) \cdot r_i(x,x_1,\dots, x_d ) \cdot X_i(x; x_1, \dots, x_d). \label{cmAZansatz}
\end{align}
%----------------------------------------------------------------------------------------------------
Then it turns out that for $L$ chosen sufficiently large\footnote{There exist upper bounds for a particular input. But usually, these bounds are too high and one tries smaller values.} there exist polynomials
%----------------------------------------------------------------------------------------------------
$X_i(x; x_1, \dots , x_d)\in \K[x][x_1, \dots, x_d]$ with $1\leq i\leq L$ and polynomials $e_i(x)\in\K[x]$ (not all zero) such that~\eqref{cmAZdiff} holds. Motivated by this fact, one searches for these unknowns $X_i$ and $e_i$ as follows.
Note that the ansatz~(\ref{cmAZdiff}) is equivalent to (see~\cite{AlmZeil})
%----------------------------------------------------------------------------------------------------
\begin{align}
&\sum_{i=1}^d [D_{x_i}r_i(x,x_1, \dots, x_d)+q_i(x,x_1, \dots, x_d)] \cdot X_i(x;x_1, \dots, x_d)\nonumber\\
&\hspace{1cm}+r_i(x,x_1, \dots, x_d)\cdot D_{x_i}X_i(x,x_1, \dots, x_d)=h(x,x_1,\ldots,x_d). \label{cmAZdiff2}
\end{align}
%----------------------------------------------------------------------------------------------------
Finally, we choose appropriate degree bounds w.r.t.\ the $x_1,\dots,x_d$ for the $X_i$ ($1\leq i\leq d$) and plug the polynomials 
with unknown coefficients from $\K[x]$ (from $\K(x)$) into~\eqref{cmAZdiff2}. By coefficient comparison this yields 
a linear system in $\K(x)$ with the unknowns $e_i(x)$ and the unknown coefficients of the polynomials $X_i$. Finally, we 
can seek a non-trivial solution for~\eqref{cmAZdiff2} and thus for~\eqref{cmAZdiff}. In the end, we clear 
denominators in $x$ such that the $e_i(x)$ turn to polynomials.\\
If $F(x;\dots,x_{i-1},u_i,x_{i+1},\dots)=0$ and $F(x;\dots,x_{i-1},o_i,x_{i+1},\dots)=0$ 
%(and hence $G(n;\dots,x_{i-1},u_i,x_{i+1},\dots)=0$ and $G(n;\dots,x_{i-1},o_i,x_{i+1},\dots)=0$ )
then
%-------------------------------------------------------------------------------------------------------
$$
{\cal I}(x):=\int_{u_d}^{o_d} \dots\int_{u_1}^{o_1}F(x;x_1, \dots, x_d) dx_1 \dots dx_d,
$$
%----------------------------------------------------------------------------------------------------
satisfies the homogeneous linear differential equation with polynomial coefficients 
%----------------------------------------------------------------------------------------------------
\begin{align}\label{cAZlindiff}
\sum_{i=0}^L e_i(x) D_{x}^i {\cal I}(x)= 0.
\end{align}
The general method now is straightforward: Given an integrand of the form~(\ref{cmAZintegrand}), we can set $L=0,$ look for degree bounds for $X_i(x,x_1, \dots , x_d)$ and try to find a 
solution of~(\ref{cAZlindiff}) by coefficient comparison. If we do not find a solution of~(\ref{cAZlindiff}) with not all $e_i(x)$'s equal to zero (with homomorphic image testng to decide non-existence efficiently), we increase $L$ by one, look for new degree bounds 
for $X_i(x,x_1, \dots , x_d)$ and try again to find a solution of~(\ref{cAZlindiff}). Again, if we do not find a solution  with not all $e_i(x)$'s equal to zero, we increase $L$ by one and repeat 
the process.
\begin{svgraybox}
The continuous Almkvist-Zeilberger algorithm is implemented in the command \texttt{cmAZ} of \MultiIntegrateP.
\end{svgraybox}
Once we found a differential equation we can make use of the differential equation solver implemented in~\HarmonicSumsP. This solver finds all solutions of holonomic differential equations that can be expressed in
terms of iterated integrals over hyperexponential alphabets~\cite{InvMellin,Abramov1994,Abramov1996,Bronstein,Petkovsek1992} (with harmonic polylogarithms~\cite{Remiddi:1999ew}, cyclotomic polylogarithms~\cite{Ablinger:2011te} and iterated integrals over root-valued alphabets~\cite{Ablinger:2014bra} as special cases); these solutions
are called d'Alembertian solutions~\cite{Abramov1994}, in addition for differential equations of order two it finds all solutions that are Liouvillian~\cite{InvMellinKovacic,Kovacic}.
\begin{example}[cmAZ]
\noindent The following problem, which was already solved in \cite{Broadhurst}, was communicated to us by D. Broadhurst. The goal is to find a differential equation satisfied by
\begin{align}\label{broadhurst}
 Y(h)=\int _0^1\int _u^1\frac{1}{\sqrt{u v (1-u) (1-v) (1-u h) (1-(1-v) h)}}dvdu.
\end{align}
In order to fit \eqref{broadhurst} to the requirements of the AZ-algorithm we transform it using the substitution $v\to u/(1 + (u-1) z),$ which leads to 
\begin{align}
 \int _0^1\int _0^1\frac{1}{\sqrt{(1-h u) (z-1) (1+(u-1) z) (h (u-1) (z-1)+z-u z-1)}}dzdu.
\end{align}
Now we can apply our implementation:
\begin{mma}
\In {\text{cmAZ}\Biggl[\frac{1}{\sqrt{(1-h u) (z-1) (1+(u-1) z) (h (u-1) (z-1)+z-u z-1)}},h,\{u,z\},\\ 
\text{AddFactors}\to \left\{(1-u)^3 (1-z)^6,(1-z)^3\right\}\Biggr]}\\
\Out {-1+2 h+2 \left(1-7 h+7 h^2\right) D_h+6 (-1+h) h (-1+2 h) D_h^2+2 (-1+h)^2 h^2 D_h^3}\\
\end{mma}
\noindent Note that in this example the integrand is not vanishing at the integration bounds, still we could derive a homogeneous differential equation, for details we refer to the next session. However, here the right hand side can be computed easily and we find the following differential equation, which is equivalent to the one found in \cite{Broadhurst}:
$$
\biggl((h-1)^2 h^2 D_h^3+3 (h-1) h (2 h-1) D_h^2+(1-7 h+7 h^2) D_h+h-\frac{1}{2}\biggr)Y(h)=\frac{\left(h^2+4 h-4\right)}{\sqrt{1-h} (2-h)^2}.
$$
In a similar way this was already proven by D. van Straten.
\end{example}
\subsection{Dealing with non-standard boundary conditions}
Unfortunately, in many cases the integrand (\ref{cmAZintegrand}) does not vanish at the integration bounds and we end up in a linear differential equation with a non-trivial inhomogeneous part which can be written as a linear combination of integrals with at least one integral operator less.
In the following we will deal with non-standard boundary conditions in two different ways, similar to the discrete case of Section \ref{Sec:AZ}.
\subsubsection{Dealing with inhomogeneous differential equations}
\label{chomansatz}
In the previous section a method that deals with the inhomogeneous recurrences was stated, here we will use similar considerations that will give rise to a recursive method. To be more precise, we consider the integral
$$
{\cal I}(x):=\int_{u_d}^{o_d} \cdots\int_{u_1}^{o_1}F(x;x_1, \dots, x_d) dx_1 \dots dx_d.
$$
Suppose that we found
\begin{equation}
\sum_{i=0}^L e_i(x) D^i_x F(x;x_1, \dots, x_d)= \sum_{i=1}^d D_{x_i} G_i(x;x_1, \dots, x_d)
\end{equation}
where at least one $G_i(x;x_1, \dots, x_d)$ does not vanish at the integration limits. By integration with respect to $x_1,\ldots,x_d$ we can deduce that ${\cal I}(x)$ satisfies the inhomogeneous linear differential 
equation
\begin{align*}
&\sum_{i=0}^L e_i(x) D^i_x {\cal I}(x)=\\
&\hspace{1cm}\sum_{i=1}^d \int_{u_d}^{o_d} \cdots \int_{u_{i-1}}^{o_{i-1}}\int_{u_{i+1}}^{o_{i+1}}\cdots \int_{u_1}^{o_1}O_i(x)dx_1\dots dx_{i-1}dx_{i+1}\dots dx_d\\
&\hspace{1cm}-\sum_{i=1}^d \int_{u_d}^{o_d} \cdots \int_{u_{i-1}}^{o_{i-1}}\int_{u_{i+1}}^{o_{i+1}}\cdots \int_{u_1}^{o_1}U_i(x)dx_1\dots dx_{i-1}dx_{i+1}\dots dx_d
\end{align*}
with
\begin{align*}
U_i(x)&:=G_i(x;x_1,\dots,x_{i-1},o_i,x_{i+1} \dots, x_d)\\
O_i(x)&:=G_i(x;x_1,\dots,x_{i-1},u_i,x_{i+1} \dots, x_d).
\end{align*}
Note that the inhomogeneous part of the above differential equation is a sum of $2\cdot d$ integrals of dimension $d-1,$ which fit again into the input class of the continuous multiple Almkvist-Zeilberger algorithm. 
Hence we can apply the algorithms to the $2\cdot d$ integrals recursively until we arrive at the base case of one-dimensional integrals for which we have to solve an inhomogeneous differential equation 
where the inhomogeneous part is free of integrals. Given the solutions for the one-dimensional integrals we can step by step find the solutions of higher dimensional integrals until we finally find the 
solution for ${\cal I}(x)$ by solving again an inhomogeneous linear differential equation and combining it with the initial conditions. Note that we have to calculate the initial conditions with respect to $x$ for all the integrals arising in this process.\\
Summarizing, we use the following strategy (note that we assume that we are able to compute the initial conditions for the arising integrals):
\begin{programcode}{Divide and conquer strategy}
\vspace{-0.6cm}
\begin{enumerate}
\item BASE CASE: If ${\cal I}(x)$ has no integration quantifiers, return ${\cal I}(x).$
\item DIVIDE: As worked out above, compute a differential equation
\begin{equation}\label{Equ:IntDiff}
a_0(x){\cal I}(x)+a_1(x)D_x{\cal I}(x)+\dots+a_d(x)D^d_x{\cal I}(x)=h(x)
\end{equation}
with polynomial coefficients
$a_i(x)\in\K[x]$, $a_m(x)\neq0$ and the right side $h(x)$ containing a linear
combination of hyperexponential multi-integrals each with less than $d$ integration
quantifiers.
\item CONQUER: Apply the strategy recursively to the simpler integrals in
$h(x)$. This results in an iterated integral expressions $\tilde{h}(x)$ with 
\begin{equation}\label{Equ:hdiffsol}
\tilde{h}(x)=h(x).
\end{equation}
If the method fails to find the $\tilde{h}(x)$ in terms of iterated integral expressions, STOP.
\item COMBINE: Given~\eqref{Equ:IntDiff} with~\eqref{Equ:hdiffsol},
compute, if possible, $\tilde{{\cal I}}(x)$ in terms of iterated integral  expressions such that
\begin{equation}
\tilde{{\cal I}}(x)={\cal I}(x)
\end{equation}
by solving the differential equation.
\end{enumerate}
\end{programcode}
\begin{svgraybox}
 This divide and conquer strategy is implemented in the command \texttt{cmAZ\-Inte\-grate} of \MultiIntegrateP.
\end{svgraybox}
\begin{remark}
We remark that this approach works nicely, if the initial conditions of the integrals in 
the inhomogeneous part can be calculated efficiently.
\end{remark}
\begin{example}[cmAZIntegrate]
We consider the integral
\begin{align}\label{cmAZExampleIntegral}
\int_{-1}^1\int_{-1}^1\int_{-1}^1\int_{-1}^1 e^{-x (w_1 w_2 + w_3 w_4)}dw_4dw_3dw_2dw_1:
\end{align}
\begin{mma}
\In {\text{\bf cmAZIntegrate}[e^{-x (w_1 w_2 + w_3 w_4)},x,\{\{w_1,-1,1\},\{w_2,-1,1\},\{w_3,-1,1\},\{w_4,-1,1\}\}]}\\
\Out {\frac{8 \left(\text{G}\left(\frac{e^{-\tau }}{\tau },\frac{e^{-\tau }}{\tau };x\right)-\text{G}\left(\frac{e^{-\tau }}{\tau },\frac{e^{\tau }}{\tau };x\right)-\text{G}\left(\frac{e^{\tau }}{\tau },\frac{e^{-\tau }}{\tau
   };x\right)+\text{G}\left(\frac{e^{\tau }}{\tau },\frac{e^{\tau }}{\tau };x\right)\right)}{x^2}}\\
\end{mma}
\noindent Note that the iterated integrals are defined recursively by
$$
\text{G}\left(f_1(\tau),f_2(\tau),\cdots,f_k(\tau);x\right)=\int_0^xf_1(\tau_1)\text{G}\left(f_2(\tau),\cdots,f_k(\tau);\tau_1\right)d\tau_1,
$$
with the special case $\text{G}(x)=1$, compare $\eg$ \cite{PochhammerSums}.\\
Here, in a first step the differential equation
\begin{align*}
&2 f(x)+x D_x f(x)=\\
& \int _{-1}^1\int _{-1}^1\int _{-1}^1 e^{x \left(-w_1 w_2+w_4\right)}dw_4dw_2dw_1
+\int _{-1}^1\int _{-1}^1\int _{-1}^1 e^{-x \left(w_1 w_2+w_4\right)}dw_4dw_2dw_1\\
&+\int _{-1}^1\int _{-1}^1\int _{-1}^1 e^{-x \left(w_2+w_3 w_4\right)}dw_4dw_3dw_2
+\int _{-1}^1\int _{-1}^1\int _{-1}^1 e^{x \left(w_2-w_3 w_4\right)}dw_4dw_3dw_2
\end{align*}
is computed. The procedure is applied recursively to all the integrals on the right hand side, which leads to
\begin{align*}
2 f(x)+x f'(x)=-\frac{8 e^{-x} \left(e^{2 x}-1\right) \left(\text{G}\left(\frac{e^{-\tau }}{\tau };x\right)-\text{G}\left(\frac{e^{\tau }}{\tau };x\right)\right)}{x^2}
\end{align*}
Finally, solving this differential equation and combining with initial conditions yields the result.
\end{example}
\subsubsection{Adapting the ansatz to find homogeneous differential equations}
\label{cinhomansatz}
In order to avoid the difficulties of inhomogeneous differential equations we adapt the ansatz.
Namely, we can always obtain a homogeneous differential equation of the form~\eqref{cAZlindiff}
by changing (\ref{cmAZansatz}) to
%----------------------------------------------------------------------------------------------------
\begin{align}
&G_i(x;x_1, \dots, x_d)=\nonumber\\&\hspace{1cm}\overline{H}(x;x_1, \dots, x_d) \cdot r_i(x_1,\dots, x_d ) \cdot X_i(x_1, \dots, x_d)(x_i-u_i)(x_i-o_i), \label{cmAZansatz2}
\end{align}
%-------------------------------------------------------------------------------------------------------
\ie the $G_i$ are forced to vanish at the integration bounds. 
Then with this ansatz (\ref{cmAZdiff2}) the underlying linear system turns into
%-------------------------------------------------------------------------------------------------------
\begin{align}\label{cmAZrec3}
&\sum_{i=1}^d [D_{x_i}r_i(x_1, \dots, x_d)+q_i(x_1, \dots, x_d)] \cdot X_i(x_1, \dots, x_d)(x_i-u_i)(x_i-o_i) \\
&\hspace{1cm}+r_i(x,x_1, \dots, x_d)\cdot D_{x_i}X_i(x_1, \dots, x_d)(x_i-u_i)(x_i-o_i)=h(x,x_1,\ldots,x_d).\nonumber
\end{align}
%--------------------------------------------------------------------------------------------------------
The general method now is straightforward: Given an integrand of the form~(\ref{cmAZintegrand}), we can set $L=0,$ look for degree bounds for $X_i(x,x_1, \dots , x_d)$ and try to find a 
solution of~(\ref{cmAZrec3}) by coefficient comparison. If we do not find a solution of~(\ref{cmAZrec3}) with not all $e_i(x)$'s equal to zero (again homomorphic image testing is used for speedups), we increase $L$ by one, look for new degree bounds 
for $X_i(x,x_1, \dots , x_d)$ and try again to find a solution of~(\ref{cmAZrec3}). Again, if we do not find a solution  with not all $e_i(x)$'s equal to zero, we increase $L$ by one and repeat 
the process.\\
Once we found a differential equation we can use the differential equation solver implemented in the package~\HarmonicSumsP\ to try to find a closed form solution.
\begin{svgraybox}
 This strategy implemented in the command \texttt{cmAZ\-Direct\-Inte\-grate} of \MultiIntegrateP.
\end{svgraybox}
\begin{remark}
The advantage of this approach is, that we do not have to deal with integrals (and initial conditions) recursively, since the differential equation is homogenous, however the additional conditions on the ansatz might increase the order of the differential equation drastically.
\end{remark}
\begin{example}[cmAZDirectIntegrate]
We consider again the integral given in \eqref{cmAZExampleIntegral}:
\begin{mma}
\In {\text{\bf cmAZDirectIntegrate}[e^{-x (w_1*w_2 + w_3*w_4)},x,\{\{w_1,-1,1\},\{w_2,-1,1\},\{w_3,-1,1\},\\ 
    \{w_4,-1,1\}\}]}\\
\Out \label{cmAZresult} {\frac{8 \left(\text{G}\left(\frac{e^{-\tau }}{\tau },\frac{e^{-\tau }}{\tau };x\right)-\text{G}\left(\frac{e^{-\tau }}{\tau },\frac{e^{\tau }}{\tau };x\right)-\text{G}\left(\frac{e^{\tau }}{\tau },\frac{e^{-\tau }}{\tau
   };x\right)+\text{G}\left(\frac{e^{\tau }}{\tau },\frac{e^{\tau }}{\tau };x\right)\right)}{x^2}}\\
\end{mma}
\noindent Here the differential equation
\begin{align*}
0&=32 \left(4 x-16 x^3+9 x^5\right) f(x)-4 \left(27-148 x^2+598 x^4-63 x^6\right) f'(x)\\
&-4 \left(117 x-568 x^3+556 x^5-9 x^7\right) f''(x)-\left(478 x^2-2919 x^4+603 x^6\right) f^{(3)}(x)\\
&-5 \left(34 x^3-247 x^5+9 x^7\right)f^{(4)}(x)-\left(23 x^4-189 x^6\right) f^{(5)}(x)-\left(x^5-9 x^7\right) f^{(6)}(x)
\end{align*}
is derived. Solving and combing it with the initial condition yields the result given in \texttt{Out[\ref{cmAZresult}]}.
\end{example}
\subsection{Computing series expansions of the integrals}
Due to time and memory limitations, not finding all solutions of the differential equations or due to missing initial conditions (in full generality) we might fail to process certain integrals using the methods described in the previous subsection. Therefore, inspired by the previous section we are seeking a method which computes $\ep$-expansions of integrals of the form (\ref{cAZhypexpint}).\\
Again we assume that the integral ${\cal I}(\ep,x)$ from (\ref{cAZhypexpint}) has a Laurent expansion in $\ep$ for $x\in\R$ with $x_\alpha<x<x_\beta$ for some $x_\alpha<0,x_\beta>0\in\R$ and 
thus it is an analytic function in $\ep$ throughout an annular region centered by $0$ where the pole at $\ep=0$ has some order~$K\in \Z$. Hence we can write it in the form
\begin{equation}
 {\cal I}(\ep,x) = \sum_{k=-K}^{\infty} \ep^k I_k(x).
\end{equation}
In the following we try to find the first coefficients $I_t(x),I_{t+1}(x),\ldots,I_{u}(x)$ in terms of iterated integral expressions of the expansion
\begin{equation}\label{expansatz}
{\cal I}(\ep, x) = I_t(x)\ep^t+I_{t+1}(x)\ep^{t+1}+I_{t+2}(x)\ep^{t+2}+\dots
\end{equation}
with $t=-K \in \Z$.
Assume that we managed to compute a differential equation satisfied by  ${\cal I}(\ep,x)$ in the form
\begin{multline}\label{expdiff}
a_0(\ep,x)J(\ep,x)+a_1(\ep,x)D_x J(\ep,x)+\dots+a_d(\ep,x)D_x^d J(\ep,x)\\
=h_{-K}(x)\ep^{-K}+h_{-K+1}(x)\ep^{-K+1}+\dots+h_{u}(x)\ep^u+\dots.
\end{multline}
In order to find such a differential equation we can use the methods presented in the previous subsections.
In the package \HarmonicSumsP\ we implemented an algorithm that tries to find \eqref{expansatz}, given a differential equation \eqref{expdiff} and suitable initial conditions given as power series expansions about $x=0$ starting from some $s\in\Z$:
\begin{equation}\label{init}
\begin{split}
 I_t(x)&=I_{t,s} x^s+I_{t,s+1} x^{s+1}+\cdots+I_{t,s+2} x^{s+d-1}+O(x^{s+d})\\
 I_{t+1}(x)&=I_{t+1,s} x^s+I_{t+1,s+1} x^{s+1}+\cdots+I_{t+1,s+2} x^{s+d-1}+O(x^{s+d})\\
 &\vdots\\
 I_{u}(x)&=I_{u,s} x^s+I_{u,s+1} x^{s+1}+\cdots+I_{u,s+2} x^{s+d-1}+O(x^{s+d})
\end{split}
\end{equation}
In the following we will illustrate the basic calculation steps of this algorithm, which can be considered as the continuous version of the algorithm presented in \cite{Bluemlein2011}; see Section \ref{Sec:AZseriesexp}.
Inserting the ansatz~\eqref{expansatz} into~\eqref{expdiff} yields
%----------------------------------------------------------------------------------------------------
\begin{equation}\label{Equ:EpDiffAnsatz}
\begin{split}
a_0(\ep,x)&\Big[I_{t}(x)\ep^{t}+I_{t+1}(x)\ep^{t+1}+I_{t+2}(x)\ep^{t+2}+\dots\Big]+\\
a_1(\ep,x)&\Big[D_x I_{t}(x)\ep^{t}+D_x I_{t+1}(x)\ep^{t+1}+D_x I_{t+2}(x)\ep^{t+2}+\dots\Big]\\
  +\dots+\\
a_d(\ep,x)&\Big[D_x^d I_{t}(x)\ep^{t}+D_x^d I_{t+1}(x)\ep^{t+1}+D_x^d I_{t+2}(x)\ep^{t+2}+\dots\Big]\\
&=h_{t}(x)\ep^t+h_{t+1}(x)\ep^{t+1}+\dots+h_{u}(x)\ep^u+\dots.
\end{split}
\end{equation}
%----------------------------------------------------------------------------------------------------
Since two Laurent series agree if they agree coefficient-wise, we obtain the following constraint for $I_{t}(x)$ by coefficient 
comparison: 
%----------------------------------------------------------------------------------------------------
\begin{equation}
\sum_{k=0}^d a_k(0,x) D_x^k I_{t}(x) = h_{t}(x),
\end{equation}
%----------------------------------------------------------------------------------------------------
with the initial condition given in~\eqref{init}.
%----------------------------------------------------------------------------------------------------
We are now in the position to try to find an explicit representation using \HarmonicSumsP's differential equation solver~\cite{InvMellin,InvMellinKovacic}. 
%----------------------------------------------------------------------------------------------------
We assume that we could find an iterated integral representation $\tilde{I}_t(x)$ such that $\tilde{I}_t(x)=I_t(x)$ for all $x\in(x_\alpha,x_\beta)$.
%------------------------------------------------------------------------------------
In order to obtain the next coefficient of the Laurent series in $\ep$, we insert $\tilde{I}_t(x)$ 
into~\eqref{Equ:EpDiffAnsatz}, which yields
%----------------------------------------------------------------------------------------------------
\begin{equation}\label{Equ:EpDiffAnsatz2}
\begin{split}
a_0(\ep,x)&\Big[I_{t+1}(x)\ep^{t+1}+I_{t+2}(x)\ep^{t+2}+I_{t+3}(x)\ep^{t+3}+\dots\Big]+\\
a_1(\ep,x)&\Big[D_x I_{t+1}(x)\ep^{t+1}+D_x I_{t+2}(x)\ep^{t+2}+D_x I_{t+3}(x)\ep^{t+3}+\dots\Big]\\
  +\dots+\\
a_d(\ep,x)&\Big[D_x^d I_{t+1}(x)\ep^{t+1}+D_x^d I_{t+2}(x)\ep^{t+2}+D_x^d I_{t+3}(x)\ep^{t+3}+\dots\Big]\\
&=\tilde{h}_{t+1}(x)\ep^{t+1}+\tilde{h}_{t+2}(x)\ep^{t+1}+\dots+\tilde{h}_{u}(x)\ep^u+\dots
\end{split}
\end{equation}
with $\tilde{h}_{i}(x)=h_{i}(x)-g_{i}(x)$, where the $g_i(x)$ satisfy 
\begin{multline*}
 a_0(\ep,x)\tilde{I}_{t}(x)+a_1(\ep,x)D_x \tilde{I}_{t}(x)+\dots+a_d(\ep,x)D_x^d \tilde{I}_{t}(x)\\=h_{t}(x)\ep^t+g_{t+1}(x)\ep^{t+1}+\dots+g_{u}(x)\ep^u+\dots.
\end{multline*}
Now we repeat the above procedure: by coefficient comparison we obtain the following constraint for $I_{t+1}(x)$:
\begin{equation}\label{step2constraint}
\sum_{k=0}^d a_k(0,x) D_x^k I_{t+1}(x) = \tilde{h}_{t+1}(x).
\end{equation}
Assuming that we can find a solution $\tilde{I}_{t+1}(x)$  of \eqref{step2constraint} in terms of iterated integrals that satisfy the initial condition from \eqref{init} such that $\tilde{I}_{t+1}(x)=I_{t+1}(x)$ for all $x\in(x_\alpha,x_\beta)$ we can update the ansatz \eqref{Equ:EpDiffAnsatz2}:
\begin{equation}\label{Equ:EpDiffAnsatz3}
\begin{split}
a_0(\ep,x)&\Big[I_{t+2}(x)\ep^{t+2}+I_{t+3}(x)\ep^{t+3}+I_{t+4}(x)\ep^{t+4}+\dots\Big]+\\
a_1(\ep,x)&\Big[D_x I_{t+2}(x)\ep^{t+2}+D_x I_{t+3}(x)\ep^{t+3}+D_x I_{t+4}(x)\ep^{t+4}+\dots\Big]\\
  +\dots+\\
a_d(\ep,x)&\Big[D_x^d I_{t+2}(x)\ep^{t+2}+D_x^d I_{t+3}(x)\ep^{t+3}+D_x^d I_{t+4}(x)\ep^{t+4}+\dots\Big]\\
&=\tilde{\tilde{h}}_{t+2}(x)\ep^{t+2}+\tilde{\tilde{h}}_{t+3}(x)\ep^{t+3}+\dots+\tilde{\tilde{h}}_{u}(x)\ep^u+\dots
\end{split}
\end{equation}
with $\tilde{\tilde{h}}_{i}(x)=\tilde{h}_{i}(x)-\tilde{g}_{i}(x)$, where the $\tilde{g}_i(x)$ satisfy 
\begin{multline*}
 a_0(\ep,x)\tilde{I}_{t+1}(x)+a_1(\ep,x)D_x \tilde{I}_{t+1}(x)+\dots+a_d(\ep,x)D_x^d \tilde{I}_{t+1}(x)\\=\tilde{h}_{t+1}(x)\ep^t+\tilde{g}_{t+2}(x)\ep^{t+2}+\dots+\tilde{g}_{u}(x)\ep^u+\dots.
\end{multline*}
We can repeat this process as long as we can compute solutions and as long as needed. The illustrated calculation steps can be summarized with the following theorem.
\medskip
%----------------------------------------------------------------------------------------------------
\begin{theorem}
Suppose we are given a linear differential equation
\begin{multline*}
a_0(\ep,x)J(\ep,x)+a_1(\ep,x)D_x J(\ep,x)+\dots+a_d(\ep,x)D_x^d J(\ep,x)\\
=h_{k}(x)\ep^k+h_{k+1}(x)\ep^{k+1}+\dots+h_{u}(x)\ep^u+\dots
\end{multline*}
of order $d$ where the $a_i(\ep,x)$ are polynomials in $x$ and $\ep$ and where the inhomogeneous part can be expanded in $\ep$ up to order $u$ in terms of expressions in iterated integrals over hyperexponential alphabets.
Consider a function which has a Laurent series expansion
\begin{equation*}
J(\ep,x)=F_{k}(x)\ep^k+F_{k+1}(x)\ep^{k+1}+\dots
\end{equation*}
and which is a solution of the given differential equation for all $x\in\R$ with $x_\alpha<x<x_\beta$ for some $x_\alpha<0,x_\beta>0\in\R$. Then together with the initial conditions  
\begin{equation*}
 F_j(x)=F_{j,s} x^s+F_{j,s+1} x^{s+1}+\cdots+F_{j,s+2} x^{s+d-1}+O(x^{s+d})
\end{equation*}
with $k\leq j\leq u$, all $F_k(x),\dots,F_u(x)$ with $x_\alpha<x<x_\beta$ can be computed in terms of expressions in iterated integrals over hyperexponential alphabets provided that the values 
$h_j(x)$ for all $j$ with $k\leq i\leq u$ and $x_\alpha<x<x_\beta$  can be computed in terms of expressions in iterated integrals over hyperexponential alphabets.
\end{theorem}
%----------------------------------------------------------------------------------------------------
This algorithm is implemented in the package \HarmonicSumsP\ and with this implementation in hand we can try to find Laurent series solutions of integrals of the form~\eqref{cAZhypexpint}.
Let ${\cal I}(\ep,n)$ be a multi-integral of the form~\eqref{cAZhypexpint} and assume that 
${\cal I}(\ep,x)$ has a series expansion~\eqref{expansatz} for all $x\in\R$ with $x_\alpha<x<x_\beta$ for some $x_\alpha<0,x_\beta>0\in\R$. If we succeed in finding a homogeneous differential equation, for instance by using the method form Section~\ref{chomansatz} we can directly apply the Laurent series differential equation solver, supposing that we can handle the initial conditions. 
\begin{svgraybox}
This strategy is implemented in the command \texttt{cmAZ\-Expanded\-Direct\-Inte\-grate} of \MultiIntegrateP.
\end{svgraybox}

\begin{example}[cmAZExpandedDirectIntegrate]
We consider the integral
\begin{align}\label{cmAZExpandedExample}
I(\ep,w)=\int_0^1\int_0^1\int_0^1e^{x y w} ((1-w) x (1-y))^{\frac{\ep}{2}} ((1-w)y(1-x) z (1-z))dzdydx
\end{align}
with the given initial condition
\begin{align*}
I(\ep,w)=&\frac{8}{3 (2+\ep )^2 (4+\ep )^2}-\frac{4 w (28+\ep  (12+\ep ))}{3 (2+\ep ) (4+\ep )^2 (6+\ep )^2}\\
&\underbrace{+\frac{w^2 (-1664+\ep  (12+\ep  (12+\ep )) (72+\ep  (16+\ep )))}{3
   (2+\ep ) (4+\ep )^2 (6+\ep )^2 (8+\ep )^2}}_{init:=}+O(w^3).
\end{align*}
We want to find the first two terms of the $\ep-$expansion of $I(\ep,w),$ \ie we want to compute $I_0(w)$ and $I_1(w)$ such that $I(\ep,w)=I_0(w)+\ep I_1(w)+O(\ep^2).$ This can be achieved by using our implementation:
\begin{mma}  
\In {\text{\bf cmAZExpandedDirectIntegrate}[I(\ep,w), w,\{\ep, 0, 1\}, \{\{x, 0, 1\}, \{y, 0, 1\}, \{z, 0, 1\}\},\\ InitValues \to init]}\\
\Out {\Biggl\{\Biggl\{\frac{1}{6}-\frac{1}{6
   w}-\frac{\text{G}\left(\frac{1-\ep^{\tau }}{\tau
   };w\right)}{6
   w^2}+\frac{\text{G}\left(\frac{1-\ep^{\tau }}{\tau
   };w\right)}{6 w},
   -\frac{1}{12 w^2}+\frac{\ep^w}{12 w^2}+\frac{1}{12
   w}-\frac{\ep^w}{12 w}\\-\frac{\text{G}\left(\frac{1}{1-\tau
   };w\right)}{12}+\frac{\text{G}\left(\frac{1}{1-\tau
   };w\right)}{12 w}-\frac{\text{G}\left(\frac{1-\ep^{\tau }}{\tau
   };w\right)}{12}
   -\frac{\text{G}\left(\frac{1-\ep^{\tau
   }}{\tau };w\right)}{12
   w^2}+\frac{\text{G}\left(\frac{1-\ep^{\tau }}{\tau
   };w\right)}{6 w}\\+\frac{\text{G}\left(\frac{1}{1-\tau
   },\frac{1-\ep^{\tau }}{\tau };w\right)}{12
   w^2}-\frac{\text{G}\left(\frac{1}{1-\tau
   },\frac{1-\ep^{\tau }}{\tau };w\right)}{12
   w}-\frac{\text{G}\left(\frac{1}{\tau
   },\frac{1-\ep^{-\tau }}{\tau };w\right)}{12
   w^2}+\frac{\text{G}\left(\frac{1}{\tau
   },\frac{1-\ep^{-\tau }}{\tau };w\right)}{12
   w}\\+\frac{\text{G}\left(\frac{1}{\tau
   },\frac{1-\ep^{\tau }}{\tau };w\right)}{12
   w^2}-\frac{\text{G}\left(\frac{1}{\tau
   },\frac{1-\ep^{\tau }}{\tau };w\right)}{12
   w}+\frac{\text{G}\left(\frac{1-\ep^{\tau }}{\tau
   },\frac{1}{1-\tau };w\right)}{12
   w^2}-\frac{\text{G}\left(\frac{1-\ep^{\tau }}{\tau
   },\frac{1}{1-\tau };w\right)}{12
   w}\\+\frac{\text{G}\left(\frac{1-\ep^{\tau }}{\tau
   },\frac{1-\ep^{-\tau }}{\tau };w\right)}{12
   w^2}-\frac{\text{G}\left(\frac{1-\ep^{\tau }}{\tau
   },\frac{1-\ep^{-\tau }}{\tau };w\right)}{12 w}\Biggr\},\{0,1\}\Biggr\}}\\
\end{mma}
\end{example}
Of course we can again think of a recursive method to compute the first coefficients, say $F_t(n),\dots,F_u(n)$ of~\eqref{expansatz}. Note that we have the same advantages and disadvantages as mentioned for the recursive method in Section \ref{cinhomansatz}, but if we assume that we can handle the initial conditions we can use to following strategy.
\begin{programcode}{Divide and conquer strategy}
\vspace{-0.6cm}
\begin{enumerate}
\item BASE CASE: If ${\cal I}(\ep, x)$ has no integration quantifiers, compute the expansion by standard methods.
\item DIVIDE: As worked out before, compute a differential equation
\begin{equation}\label{Equ:IntDiffEq}
a_0(\ep,x){\cal I}(\ep, x)+\dots+a_d(\ep,x)D_x^d{\cal I}(\ep, x)=h(\ep,x)
\end{equation}
with polynomial coefficients
$a_i(\ep,x)\in\K[\ep,x]$, $a_m(\ep,x)\neq0$ and the right side $h(\ep,x)$ containing a linear
combination of hyperexponential multi-integrals each with less than $d$ integration
quantifiers.
\item CONQUER: Apply the strategy recursively to the simpler integrals in
$h(\ep,x)$. This results in an expansion of the form
\begin{equation}\label{Equ:chExpansion}
h(\ep,
x)=h_t(x)+h_1(x)\ep+\dots+h_u(x)\ep^u+O(\ep^{u+1});
\end{equation}
if the method fails to find the $h_t(x),\dots,h_u(x)$ in terms of iterated integral expressions, STOP.
\item COMBINE: Given~\eqref{Equ:IntDiffEq} with ~\eqref{Equ:chExpansion},
compute, if possible, the $F_t(n),\dots,F_u(n)$ of~\eqref{expansatz} in terms of
iterated integral expressions by using \HarmonicSumsP.
\end{enumerate}
\end{programcode}
\begin{svgraybox}
This divide and conquer strategy is implemented in the command \texttt{cmAZ\-Expanded\-Integrate} of \MultiIntegrateP.
\end{svgraybox}
\begin{example}[cmAZExpandedIntegrate]
Again we consider the integral given in \eqref{cmAZExpandedExample} with the same initial condition. In order to compute the first two terms of the $\ep-$expansion of $I(\ep,w),$ we can also use the following function call:
\begin{mma}  
\In {\text{\bf cmAZExpandedIntegrate}[I(\ep,w), w,\{\ep, 0, 1\}, \{\{x, 0, 1\}, \{y, 0, 1\}, \{z, 0, 1\}\},\\ InitValues \to init]}\\
\Out {\Biggl\{\Biggl\{\frac{1}{6}-\frac{1}{6
   w}-\frac{\text{G}\left(\frac{1-\ep^{\tau }}{\tau
   };w\right)}{6
   w^2}+\frac{\text{G}\left(\frac{1-\ep^{\tau }}{\tau
   };w\right)}{6 w},
   -\frac{1}{12 w^2}+\frac{\ep^w}{12 w^2}+\frac{1}{12
   w}-\frac{\ep^w}{12 w}\\-\frac{\text{G}\left(\frac{1}{1-\tau
   };w\right)}{12}+\frac{\text{G}\left(\frac{1}{1-\tau
   };w\right)}{12 w}-\frac{\text{G}\left(\frac{1-\ep^{\tau }}{\tau
   };w\right)}{12}
   -\frac{\text{G}\left(\frac{1-\ep^{\tau
   }}{\tau };w\right)}{12
   w^2}+\frac{\text{G}\left(\frac{1-\ep^{\tau }}{\tau
   };w\right)}{6 w}\\+\frac{\text{G}\left(\frac{1}{1-\tau
   },\frac{1-\ep^{\tau }}{\tau };w\right)}{12
   w^2}-\frac{\text{G}\left(\frac{1}{1-\tau
   },\frac{1-\ep^{\tau }}{\tau };w\right)}{12
   w}-\frac{\text{G}\left(\frac{1}{\tau
   },\frac{1-\ep^{-\tau }}{\tau };w\right)}{12
   w^2}+\frac{\text{G}\left(\frac{1}{\tau
   },\frac{1-\ep^{-\tau }}{\tau };w\right)}{12
   w}\\+\frac{\text{G}\left(\frac{1}{\tau
   },\frac{1-\ep^{\tau }}{\tau };w\right)}{12
   w^2}-\frac{\text{G}\left(\frac{1}{\tau
   },\frac{1-\ep^{\tau }}{\tau };w\right)}{12
   w}+\frac{\text{G}\left(\frac{1-\ep^{\tau }}{\tau
   },\frac{1}{1-\tau };w\right)}{12
   w^2}-\frac{\text{G}\left(\frac{1-\ep^{\tau }}{\tau
   },\frac{1}{1-\tau };w\right)}{12
   w}\\+\frac{\text{G}\left(\frac{1-\ep^{\tau }}{\tau
   },\frac{1-\ep^{-\tau }}{\tau };w\right)}{12
   w^2}-\frac{\text{G}\left(\frac{1-\ep^{\tau }}{\tau
   },\frac{1-\ep^{-\tau }}{\tau };w\right)}{12 w}\Biggr\},\{0,1\}\Biggr\}}\\
\end{mma}
\end{example}
\section*{Conclusion}
 In this paper we summarize the theoretical background of our package \MultiIntegrateP\ which can be downloaded at \url{https://risc.jku.at/software} and which provides several methods to deal with multiple integrals over hyperexponential integrands.
\begin{acknowledgement}
This work was supported by the Austrian Science Fund (FWF) grant SFB F50 (F5009-N15) and  by the bilateral project WTZ BG 03/2019 (KP-06-Austria/8/2019), funded by  OeAD (Austria) and Bulgarian National Science Fund. The author would like to thank C.~Schneider for useful discussions.
\end{acknowledgement}

\end{document}